# Possible itinerant excitations and quantum spin state transitions in the effective spin-1/2 triangular-lattice antiferromagnet Na$_2$BaCo(PO$_4$)$_2$

N. Li[1,6], Q. Huang[2,6], X. Y. Yue[3,6], W. J. Chu[1], Q. Chen[2], E. S. Choi[4], X. Zhao[5], H. D. Zhou[2]★, and X. F. Sun[1,3]★

[1]Department of Physics, Hefei National Laboratory for Physical Sciences at Microscale, and Key Laboratory of Strongly-Coupled Quantum Matter Physics (CAS), University of Science and Technology of China, Hefei, Anhui 230026, People's Republic of China

[2]Department of Physics and Astronomy, University of Tennessee, Knoxville, Tennessee 37996-1200, USA

[3]Institute of Physical Science and Information Technology, Anhui University, Hefei, Anhui 230601, People's Republic of China

[4]National High Magnetic Field Laboratory, Florida State University, Tallahassee, FL 32310-3706, USA

[5]School of Physical Sciences, University of Science and Technology of China, Hefei, Anhui 230026, People's Republic of China

[6]These authors contributed equally: N. Li, Q. Huang, X. Y. Yue.

★email: hzhou10@utk.edu; xfsun@ustc.edu.cn

**The most fascinating feature of certain two-dimensional (2D) gapless quantum spin liquid (QSL) is that their spinon excitations behave like the fermionic carriers of a paramagnetic metal. The spinon Fermi surface is then expected to produce a linear increase of the thermal conductivity with temperature that should manifest via a residual value ($\kappa_0/T$) in the zero-temperature limit. However, this linear in $T$ behavior has been reported for very few QSL candidates. Here, we studied the ultralow-temperature thermal conductivity of an effective spin-1/2 triangular QSL candidate Na$_2$BaCo(PO$_4$)$_2$, which has an antiferromagnetic order at very low temperature ($T_N$ ~ 148 mK), and observed a finite $\kappa_0/T$ extrapolated from the data above $T_N$. Moreover, while approaching zero temperature, it exhibits series of quantum spin state transitions with applied field along the $c$ axis. These observations indicate that Na$_2$BaCo(PO$_4$)$_2$ possibly behaves as a gapless QSL with itinerant spin excitations above $T_N$ and its strong quantum spin fluctuations persist below $T_N$.**

The two-dimensional (2D) triangular lattice antiferromagnet (TAF) with spin-1/2 is one of the simplest geometrically frustrated systems with strong quantum spin fluctuations, which recently has caught attention due to its exotic quantum magnetism. One celebrated example is the 2D gapless quantum spin liquid (QSL)[1-4] that can host non-abelian quasiparticle[5] and fractional excitations[6,7] known as spinons[8,9], which allows quantum mechanical encryption and transportation of information with potential for creating a qubit that is protected against environmental influences[10]. Three experimental hallmarks have been widely accepted as evidence for spinon, including (i) a broad continuous magnetic intensity in the inelastic neutron scattering (INS) spectrum[6,11,12]; (ii) a large magnetic specific heat with power law ($C \sim T^{\alpha}$) temperature dependence[13-15]; and (iii) a non-zero residual thermal conductivity $\kappa_0/T$ in the zero-temperature limit[16-19]. While most of the suggested 2D gapless QSLs exhibit the first two hallmarks, they do not exhibit the third one. In reality, so far only the organic $EtMe_3Sb[Pd(dmit)_2]_2$ reported by Yamashita *et al.*[18,19] and the inorganic $1T-TaS_2$ reported by Murayama *et al.*[20] exhibit a non-zero $\kappa_0/T$ term, both of which are spin-1/2 TAFs. However, some other groups also reported a zero $\kappa_0/T$ term in these two materials, raising a controversy[21-23]. For another QSL candidate pyrochlore $Tb_2Ti_2O_7$, a saturated value of $\kappa/T$ at 0.3 K was reported which resembles that of a dirty metal[24]. For other oxides, such as $YbMgGaO_4$[25], another TAF with effective spin-1/2 $Yb^{3+}$ ions, and $Ca_{10}Cr_7O_{28}$ with bilayer kagome lattice[26], the reported $\kappa_0/T$ term tends to be zero upon approaching zero temperature. This behavior could be closely related to the chemical disorder in both cases. For instance, $YbMgGaO_4$ has $Mg^{2+}/Ga^{3+}$ site mixture[27] and $Ca_{10}Cr_7O_{28}$ has disorder among the two different $Cr^{3+}$ positions[28,29].

Another example of exotic magnetism in spin-1/2 TAFs is the quantum spin state transition. The theoretical studies have proposed that in a spin-1/2 TAF, the quantum spin fluctuations (QSFs) stabilize a novel up up down (UUD) phase while approaching zero temperature with the applied field parallel to either easy plane or easy axis[30,31]. This UUD phase exhibits itself as a magnetization plateau within a certain magnetic field regime and with one-third of the saturation moment. Experimentally, it is very rare to observe such a UUD phase in TAFs while approaching zero temperature. One example is $Ba_3CoSb_2O_9$, another TAF with effective spin-1/2 $Co^{2+}$ ions, which orders around 3.5 K and exhibits a UUD phase at ultralow temperatures[32-34]. More recently, the UUD phase also has been proposed for $AYbCh_2$ ($A$ = Na and Cs, $Ch$ = O, S, Se), one TAF family with effective spin-1/2 $Yb^{3+}$ ions[35-38]. Further detailed experimental and theoretical studies on $Ba_3CoSb_2O_9$ revealed more complex quantum spin state transitions (QSSTs)[39-45]. Specifically, with increasing field along the *ab* plane, its 120° spin structure at zero field is followed by a canted 120° spin structure, the UUD phase, a coplanar phase (the V phase), and another coplanar phase (the V′ phase) before entering the fully polarized state. While for *B // c*, the 120° spin structure will be followed by an umbrella spin structure and a V phase.

While searching for new spin-1/2 TAFs without chemical disorder to explore the novel physics of QSL and QSSTs, the new Co-based triangular lattice antiferromagnet $Na_2BaCo(PO_4)_2$[46] caught our attention. This system has a trigonal crystal structure with lattice parameter $a$ = 5.3185 Å and $c$ = 7.0081 Å. The magnetic $CoO_6$ octahedra form a triangular network in the *ab* plane, separated by a layer of nonmagnetic $BaO_{12}$ polyhedra. Meanwhile, the $Na^+$ ions fill the gaps in the $CoO_6$ layers (Figs. 1a and 1b). Overall, no site mixture among the ions has been observed. Due to its Kramers ion nature, the $Co^{2+}$ ions can be treated as effective spin-1/2 at low temperatures. The magnetic susceptibility, INS spectrum, and specific heat data show no magnetic ordering down to 50 mK but with large magnetic specific heat and localized low-energy spin fluctuations. Then, as discussed above, to confirm whether this system is a truly gapless QSL or not, it is crucial to look for the possible existence of itinerant spinons. Moreover, until now, no magnetic phase diagram has been reported for this new TAF and its possibility for QSSTs is awaiting exploration.

**Results**

**Magnetic susceptibility.** By following ref.[46]'s recipe, we grew single crystals of $Na_2BaCo(PO_4)_2$. Figure 1c shows the inverse of the DC magnetic susceptibility ($1/\chi$) with applied field *B // ab*. A change of slope is observed around 50 K. The effective moment is estimated to be 5.37 $\mu_B$ for 150 < *T* < 300 K and 4.0 $\mu_B$ for 2 < *T* < 20 K by using the linear Curie-Weiss fittings. This decrease of effective moment indicates a crossover of spin state for $Co^{2+}$ ions from *S* = 3/2 at high temperatures to an effective spin-1/2 at low temperatures. For $Co^{2+}$ ions in an octahedral environment, as for $Na_2BaCo(PO_4)_2$, the crystal field and spin-orbital coupling can lead to a Kramers doublet with the effective spin-1/2 as the ground state. For other triangular lattice antiferromagnets with octahedral Co sites, such as $Ba_3CoSb_2O_9$[32] and $ACoB_3$ (*A* = Cs, Rb; *B* = Cl, Br)[47], the ground state also has effective spin-1/2. Therefore, the $\theta_{CW}$ = −2.5 K from the low-temperatures fitting represents its intrinsic antiferromagnetic exchange energy. According to the mean field theory, $\theta_{CW}$ is given as $(-zJS(S+1))/3k_B$, where *J* is the exchange interaction of the Heisenberg Hamiltonian $J\sum_{(i,j)} S_i S_j$, and $z$ is the number of nearest neighbors. For the effective $S = 1/2$ triangular lattice with $z = 6$, we obtained $J/k_B = -2/3\theta_{CW} = 1.7$ K.

**Thermal conductivity.** Figure 2a shows the zero-field thermal conductivity of Na$_2$BaCo(PO$_4$)$_2$ in the temperature range of 70 mK – 30 K. At higher temperature, it behaves like a usual insulating crystal. The peak at 12 K with a large value of 90 WK$^{-1}$m$^{-1}$ can be understood as the so-called phonon peak. It is notable that such a large phonon peak indicates high quality of the single crystal sample. Also shown are the thermal conductivity in 14 T field, either along the $c$ or the $a$ axis, which can increase the $\kappa$ at most temperatures up to 7 K.

Figures 2b, 2c and 2d show the ultralow-temperature thermal conductivity at 0 T and 14 T fields. Several features are noteworthy. First, all these data are well fitted by $\kappa/T = \kappa_0/T + bT^2$ with $b$ as a constant in a very broad temperature range (from several tens to 500 mK or more, particularly up to 700 mK for zero field), while the fitting parameters $\kappa_0/T$ and $b$ are clearly different for them. Second, in zero field the fitting gives $\kappa_0/T = 0.0062$ WK$^{-2}$m$^{-1}$, that is, the presence of a residual value in $\kappa/T$ while extending to zero temperature is clearly resolved. Third, the fitting curves in Fig. 2 yield intercepts of $0 \pm 0.0005$ WK$^{-2}$m$^{-1}$ for data with 14 T // $c$ and 14 T // $a$. The error is at least one order of magnitude smaller than the zero-field $\kappa_0/T$ value, which indicates zero $\kappa_0/T$ for the 14 T data.

**Specific heat.** As shown in the inset to Fig. 2b, the zero-field $\kappa(T)$ data also shows a very weak anomaly around 100 mK. To learn the nature of this anomaly, the specific heat ($C_p$) was measured at very low temperatures down to 50 mK, as shown in Fig. 3. At zero field, the $C_p$ exhibits a broad peak around 630 mK followed by a small and sharp peak at 148 mK. This sharp peak should represent an antiferromagnetic ordering, which is likely related to the anomaly observed from the zero field $\kappa(T)$. With increasing field along the $c$ axis, this peak's position shifts to ~ 310 mK for $B = 0.5$ and 1 T; meanwhile, its intensity abruptly increases for $B = 0.5$ and 1 T and then this peak disappears for $B = 1.5$ T, as shown in Fig. 3a. By assuming that the lattice contribution can be described by $C_{ph} = \beta T^3 + \beta_5 T^5 + \beta_7 T^7$ with $\beta = 8.83 \times 10^{-4}$ JK$^{-4}$mol$^{-1}$, $\beta_5 = -3.32 \times 10^{-7}$ JK$^{-6}$mol$^{-1}$, and $\beta_7 = 6.67 \times 10^{-11}$ JK$^{-8}$mol$^{-1}$ (See Supplementary Figure 1), the change of magnetic entropy below 4 K, $\Delta S_{mag}$, was calculated by integrating $(C_p - C_{ph})/T$ (Fig. 3b). The obtained values are 5.1 and 5.4 JK$^{-1}$mol$^{-1}$ for $B = 0$ and 1 T, respectively, which are approaching the value of $R\ln2$. This is another strong evidence that Na$_2$BaCo(PO$_4$)$_2$ can be treated as an effective spin-1/2 system. At zero field, the recovered entropy below 200 mK (where the peak starts) is 1.6 JK$^{-1}$mol$^{-1}$. This is only 28% of $R\ln2$, which indicates the strong spin fluctuations above $T_N$. A small upturn of the specific heat is observed at the lowest temperatures, which could be attributed to a contribution from the nuclear entropy. Figure 3c shows the specific heat data for $B // a$. Similar to the results for $B // c$, the low magnetic fields along the $a$ axis can also change the position of the peak at 148 mK but with weaker field dependence.

**Residual thermal conductivity.** It is abnormal for Na$_2$BaCo(PO$_4$)$_2$ to exhibit a non-zero $\kappa_0/T$ term extrapolated from the data above $T_N$. One possible scenario is that it behaves as a QSL above $T_N$ with gapless magnetic excitations, which give rise to power-law temperature dependences of the low temperature physical properties. Indeed, the reported INS spectrum[46] and specific heat data reported here support the presence of strong spin fluctuations above $T_N$. While the 2D QSL is stable at zero temperature in the strict sense, it is also known that QSL behavior, such as spinon excitations can survive at a finite temperature regime if the temperature scale is smaller than the exchange interaction, $J$. To our knowledge, a couple of quantum magnets exhibit quantum spin disordered states including QSL in a temperature range $T_N < T < J$ due to the combination of strong geometrical frustration with enhanced quantum spin fluctuations for spin-1/2, as present in Na$_2$BaCo(PO$_4$)$_2$ with $T_N$ = 148 mK and $J \sim$ 1.7 K. One good example is the Volborthite, Cu$_3$V$_2$O$_7$(OH)$_2 \cdot$2H$_2$O, with a 2D distorted kagome lattice of Cu$^{2+}$ ($S = 1/2$) ions, which antiferromagnetically orders at $T_N \sim$ 1 K with exchange constant $J \sim$ 60 K[48,49]. A finite linear $T$-dependent contribution of specific heat extrapolated to $T = 0$ K[48] and a negative thermal Hall conductivity observed above $T_N$[49] both strongly support the presence of a QSL state with gapless spin excitations above $T_N$ for Volborthite. While no clear non-zero $\kappa_0/T$ term was observed for Volborthite due to its relatively high ordering temperature, the estimated mean free path of the spin excitations from the 8 K magnetic thermal conductivity is about 80 times its inter-spin distance, which indicates its spin excitations are highly mobile[49]. The related theoretical work also proposes the existence of spinon Fermi surface in Volborthite above $T_N$[50,51]. Another relevant example is pyrochlore Yb$_2$Ti$_2$O$_7$ with effective spin-1/2 Yb$^{3+}$ ions, which ferromagnetically orders at $T_C \sim$ 0.2 K[52]. For Yb$_2$Ti$_2$O$_7$, the XY and off diagonal components of the interactions, $J_\perp \sim$ 0.58

K and $J_{z\pm}$ ~ 1.7 K, respectively, produce quantum spin fluctuations[53,54]. Its reported specific data suggests strong quantum fluctuations above $T_C$[55,56]. Its observed diffuse scattering and pinch point structure of the INS spectrum and related model calculation further suggest the presence of a quantum spin ice state above $T_C$[57]. Lately, the unusual behavior of the magneto-thermal conductivity[58] and thermal Hall conductivity[59] suggests the emergence of highly itinerant quantum magnetic monopoles in this quantum spin ice state.

By following ref.18's method, we estimate the mean free path ($l_s$) and life time of the spin excitation ($\tau_s$) of the quasiparticles responsible for the excitations in $Na_2BaCo(PO_4)_2$ by calculating $\frac{\kappa_0}{T} = \frac{\pi k_B^2}{9\hbar} \frac{l_s}{ad} = \frac{\pi}{9}\left(\frac{k_B}{\hbar}\right)^2 \frac{J}{d}\tau_s$. Here, $a$ (~ 5.32 Å) and $d$ (~ 7.01 Å) are nearest-neighbor and interlayer spin distance, respectively. From the observed $\kappa_0/T = 0.0062$ WK$^{-2}$m$^{-1}$, the $l_s$ is obtained as 36.6 Å, indicating that the excitations are mobile to a distance 7 times as long as the inter-spin distance without being scattered. Third, in high magnetic field of 14 T, although the $\kappa$ is much larger than the zero-field data, the fitting gives a negligibly small value of, or vanishing $\kappa_0/T$. This is reasonable since 14 T is strong enough to polarize all spins and completely suppress the spinon excitations of the QSL state. Thus, the 14 T data should be a result of pure phonon heat transport. From the specific heat data (See Supplementary Figure 1), it is found that the phonon specific heat can be approximated as $C_{ph} = \beta T^3$ at very low temperatures with the coefficient $\beta = 8.83 \times 10^{-4}$ JK$^{-4}$mol$^{-1}$. The phonon velocity can be calculated from the $\beta$ value as $v_{ph} = 2430$ ms$^{-1}$. The phonon thermal conductivity in the ballistic scattering limit is $\kappa_{ph} = (1/3)C_{ph}v_{ph}l_{ph}$, where the phonon mean free path is determined by the averaged sample width of $l_{ph} = 2\sqrt{A/\pi} = 0.32$ mm for this sample. Thus, the phonon thermal conductivity at low temperature is expected as $\kappa_{ph} = 2.21 \times T^3$ WK$^{-1}$m$^{-1}$. Note that this estimation is different from the 14 T data by only a factor of 2, which is acceptable. If one assumes that the 14 T data is purely due to the phonon term, much smaller signal in zero field indicates that the phonon ballistic scattering limit is not achieved, although $\kappa/T$ nicely follows $\kappa_0/T + bT^2$. Therefore, in zero field the phonons are always suffering some scattering effect besides the boundary. Apparently, at very low temperatures only the magnetic excitations can take the role of phonon scattering.

Another possible scenario is that this non-zero $\kappa_0/T$ term is related to other abnormal quasiparticles besides spinon, which means the high-$T$ (> $T_N$) phase may not be ascribed to the QSL. Either way, future studies are desirable to learn the exact origin for this interesting residual thermal conductivity in Na$_2$BaCo(PO$_4$)$_2$.

**Field dependence of thermal conductivity and AC susceptibility.** The dramatic change of the $C_p$ peak with $B \mathbin{/\mkern-6mu/} c$ suggests the possibility of spin state transitions. For further investigation, more detailed $\kappa$ and AC susceptibility in magnetic fields were measured. For $B \mathbin{/\mkern-6mu/} c$, the $\kappa(B)$ curve at 92 mK exhibits four minima at $B_{c1}$, $B_{c2}$, $B_{c3}$, and $B_{c4}$ (Fig. 4a). With increasing temperatures, $\kappa(B)$ only exhibits two minima at 151 mK and no minimum at $T > 300$ mK. The $\kappa(T)$ measured at 0.5 and 1.0 T (Fig. 4b) clearly shows a slope change around 310 mK, which is consistent with the $C_p$ peaks' position measured at the same fields. The AC susceptibility, $\chi'$, measured at 22 mK (Fig. 4c) shows three peaks at $B_{c1}$, $B_{c2}$, and $B_{c3}$. The values of these three critical fields are consistent with the $B_{c1}$, $B_{c2}$, and $B_{c3}$ observed from $\kappa(B)$. With increasing temperatures, the $B_{c1}$ peak shifts to higher fields and the $B_{c2}$ and $B_{c3}$ peaks shifts to lower fields. At $T > 280$ mK, the peaks almost disappear. Since the measured $\chi'$ shows no frequency dependence (not shown here), it could be approximately treated as the derivative of the DC magnetization $M(B)$. We calculated $M(B)$ by integrating $\chi'$. The obtained $M(B)$ at 22 mK (Fig. 4d) clearly shows a plateau regime between $B_{c1}$ and $B_{c2}$ and a slope change at $B_{c3}$ followed by saturation around 2.5 T. Although we cannot infer the absolute value of $M(B)$ here, it is obvious that the magnetization of the plateau (around 0.29, here we scaled the $M$ value to the 3 T value) is around 1/3 of the saturation value (around 0.84 after we subtract the Van Vleck paramagnetic background, which is the upper dashed line in Fig. 4d).

For comparison, the above measurements were also performed for $B \mathbin{/\mkern-6mu/} a$. The $\kappa(B)$ curve (Fig. 5a) at 92 mK shows two minima at $B_{a1}$ and $B_{a2}$, while the $\kappa(T)$ (Fig. 5b) measured at different fields shows no obvious slope change. The $\chi'$ (Fig. 5c) measured at 25 mK shows a broad peak around $B_{a1}$ and a sharp peak around $B_{a2}$. With increasing temperature, $B_{a1}$ and $B_{a2}$ shift to higher and lower fields, respectively, and both disappear at $T > 290$ mK. The calculated $M(B)$ at 25 mK shows a slope change around the $1/3M_s$ position. These results are clearly different from those for $B \mathbin{/\mkern-6mu/} c$.

**Phase diagram.** Based on the critical fields and ordering temperatures presented above, the magnetic phase diagrams for $B // c$ and $B // a$ are constructed in Fig. 6. For $B // a$, since both the $\kappa(B)$ and $\chi'$ data consistently show two critical fields and $B_{a2}$ is very close to the field where the magnetization becomes flat (or saturated), it is natural for us to assign $B_{a2}$ as the saturation field and $B_{a1}$ as a critical field for a spin state transition. On the other hand, for $B // c$, the $\kappa(B)$ exhibits four critical fields while the $\chi'$ shows three. Here we assign the $B_{c4}$ as the saturation field for two reasons: (i) if we assign $B_{c3}$ as the saturation field, it will be difficult to understand why there is still a possible spin state transition at $B_{c4} > B_{c3}$ after all spins have been polarized; (ii) a close look of the calculated $M(B)$ curve shows that $B_{c3}$ represents a slope change before the magnetization becomes flat, which most likely represents a spin state transition. One possible situation is that since this $B_{c3}$ peak of $\chi'$ data is so close to the saturation field position, it may smear out the expected $\chi'$ peak at $B_{c4}$. Accordingly, besides the paramagnetic phase at high temperatures and fully polarized phase at high fields, with increasing field, there are four phases for $B // c$ (Fig. 6a) and two phases for $B // a$ (Fig. 6b).

Now we compare the phase diagrams of $Na_2BaCo(PO_4)_2$ to those of $Ba_3CoSb_2O_9$ listed in the introduction. For $B // c$ we are confident that the phase II is the UUD phase based on the $1/3 M_s$ plateau observed at 22 mK. Since the 120° spin structure is a pre-required phase for the appearance of UUD phase, we ascribe the phase I as the canted 120° spin structure. Whether the phase III and IV are the V and V' phase or the phase I and II for $B // a$ are the umbrella and V phases or not cannot be said at this stage. Further studies such as neutron diffraction are needed to address this question.

We emphasize that in $Na_2BaCo(PO_4)_2$, the UUD phase only survives for $B // c$, which strongly suggests its easy axis anisotropy as the theory predicted[31]. $Ba_3CoSb_2O_9$ and $AYbCh_2$ are both TAFs with easy plane anisotropy. To our knowledge, $Na_2BaCo(PO_4)_2$ is a very rare example of spin-1/2 TAF with single crystalline form to exhibit series of QSSTs along the easy axis. Another two examples for spin-1/2 TAFs with easy axis anisotropy to show UUD phase are $Ba_3CoNb_2O_9$[60,61] and $Ba_2La_2CoTe_2O_{12}$[62], but both of them are polycrystalline form.

**Discussion**

In summary, we clearly observed a nonzero residual thermal conductivity, $\kappa_0/T$, extrapolated from the data above $T_N$ (~ 148 mK) in an effective spin-1/2 triangular antiferromagnet $Na_2BaCo(PO_4)_2$.

This abnormal feature indicates that Na$_2$BaCo(PO$_4$)$_2$ possibly behaves as a gapless QSL with itinerant spin excitations above $T_N$. Moreover, its strong quantum spin fluctuations persist below $T_N$ and help to stabilize a series of spin state transitions while approaching zero temperature. With applied field along the $c$ axis, this includes the UUD phase with a 1/3$M_s$ magnetization plateau. This makes Na$_2$BaCo(PO$_4$)$_2$ a unique TAF with easy axis anisotropy to exhibit a UUD phase.

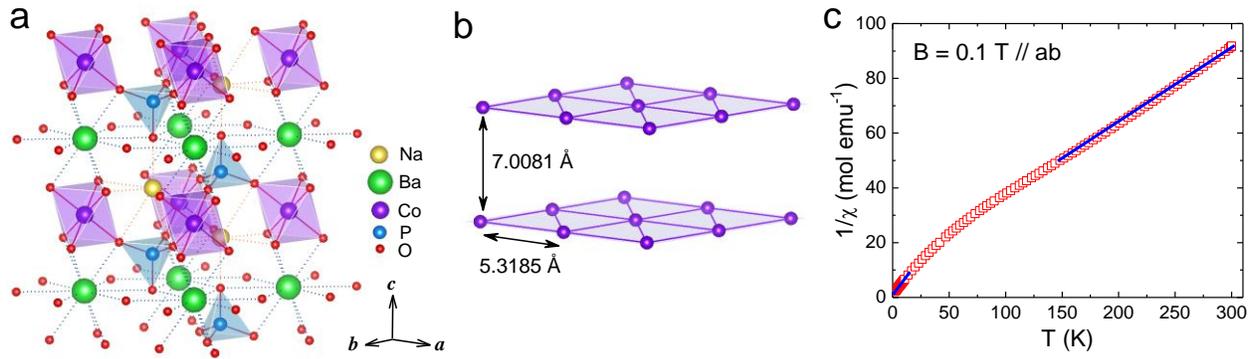

**Figure 1  Structure and magnetic susceptibility of Na$_2$BaCo(PO$_4$)$_2$. a**, The crystallographic structure. **b**, The triangular lattice of Co$^{2+}$ ions in the *ab* plane. **c**, The inverse of the DC susceptibility measured with 0.1 T magnetic field along the *ab* plane. The solid lines are the Curie-Weiss fittings to high-temperature or low-temperature data.

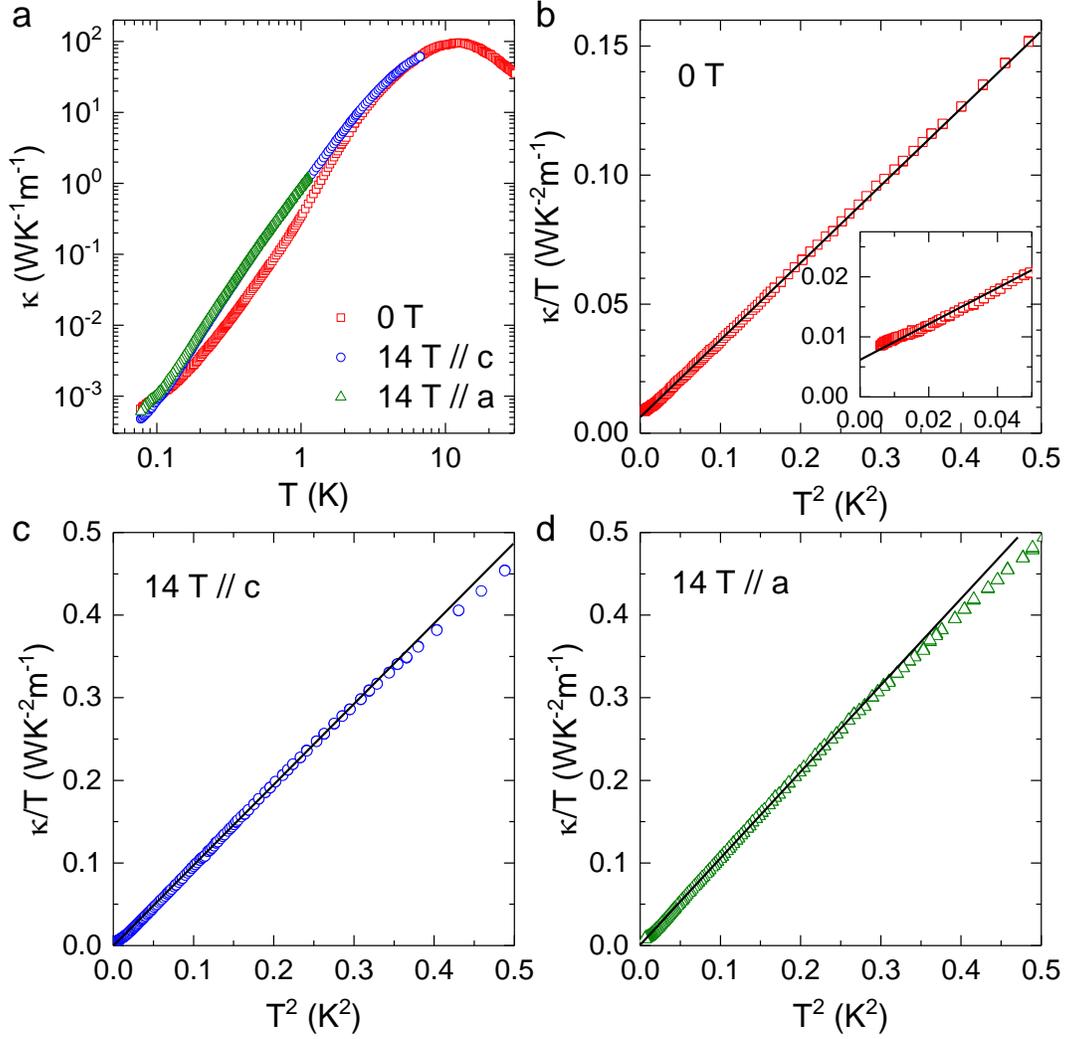

**Figure 2  Ultralow-temperature thermal conductivity of Na$_2$BaCo(PO$_4$)$_2$. a**, The zero-field thermal conductivity of Na$_2$BaCo(PO$_4$)$_2$ in temperature range of 70 mK – 30 K. The heat current is along the *a* axis. The peak at 12 K is the so-called phonon peak. Also shown are the thermal conductivity in 14 T magnetic field along the *c* or the *a* axis. In most of this temperature region, high magnetic field enhances the thermal conductivity. **b**, Data in zero field plotted in $\kappa/T$ vs $T^2$. The solid line is a linear fit of the data at $T < 700$ mK. A nonzero residual thermal conductivity $\kappa_0/T = 0.0062$ WK$^{-2}$m$^{-1}$ is resolved. The inset shows a magnified view of the lowest-temperature data. There is a very weak anomaly at $T < 100$ mK. **c,d,** Thermal conductivity in 14 T magnetic field plotted as $\kappa/T$ vs $T^2$. The solid lines are a linear fits for data at $T < 550$ mK (for *B* // *c*) and at $T < 500$ mK (for *B* // *a*). There is no residual term ($\kappa_0/T = 0$).

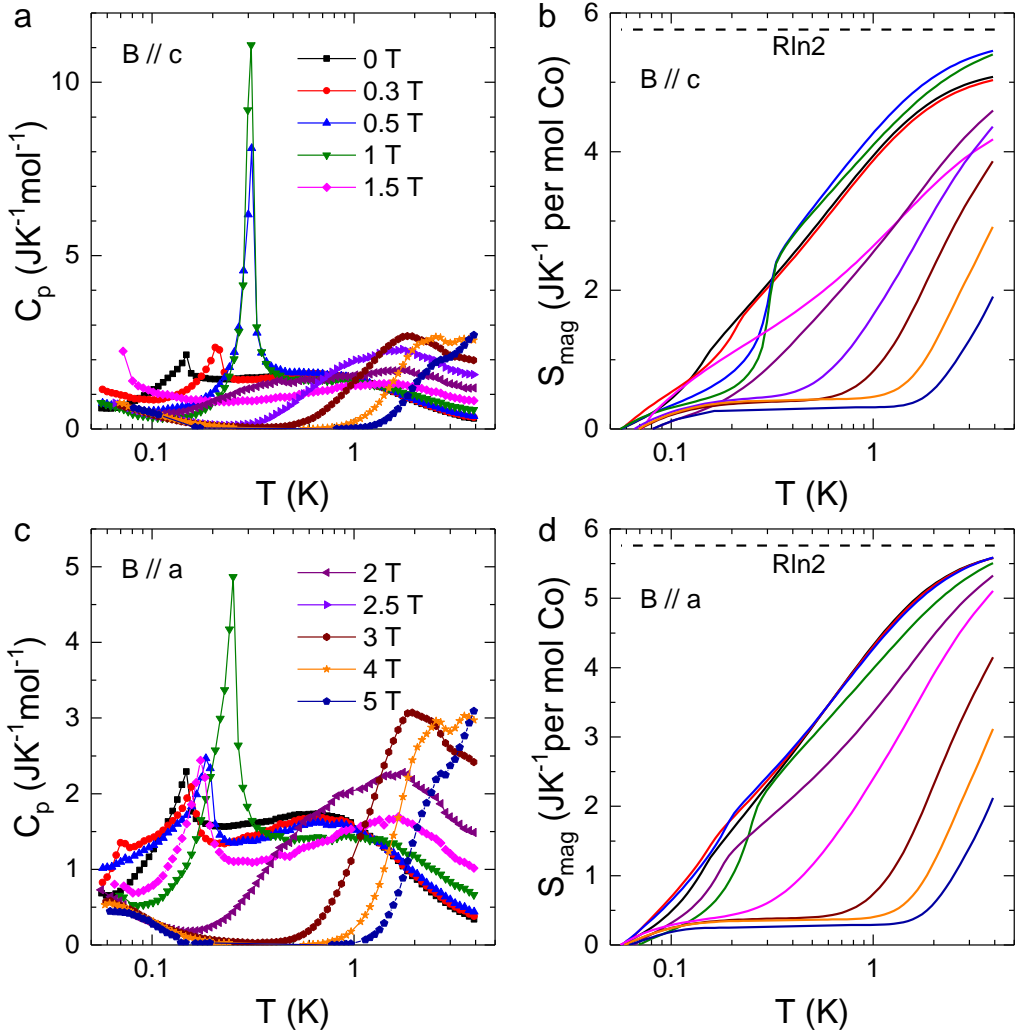

**Figure 3 Ultralow-temperature specific heat of Na$_2$BaCo(PO$_4$)$_2$. a,c,** The zero-field data and those in different magnetic fields along the *c* axis or the *a* axis. The temperature range is 50 mK – 4 K. **b,d,** The magnetic entropy for *B // c* and *B // a*, obtained by integrating the magnetic specific heat.

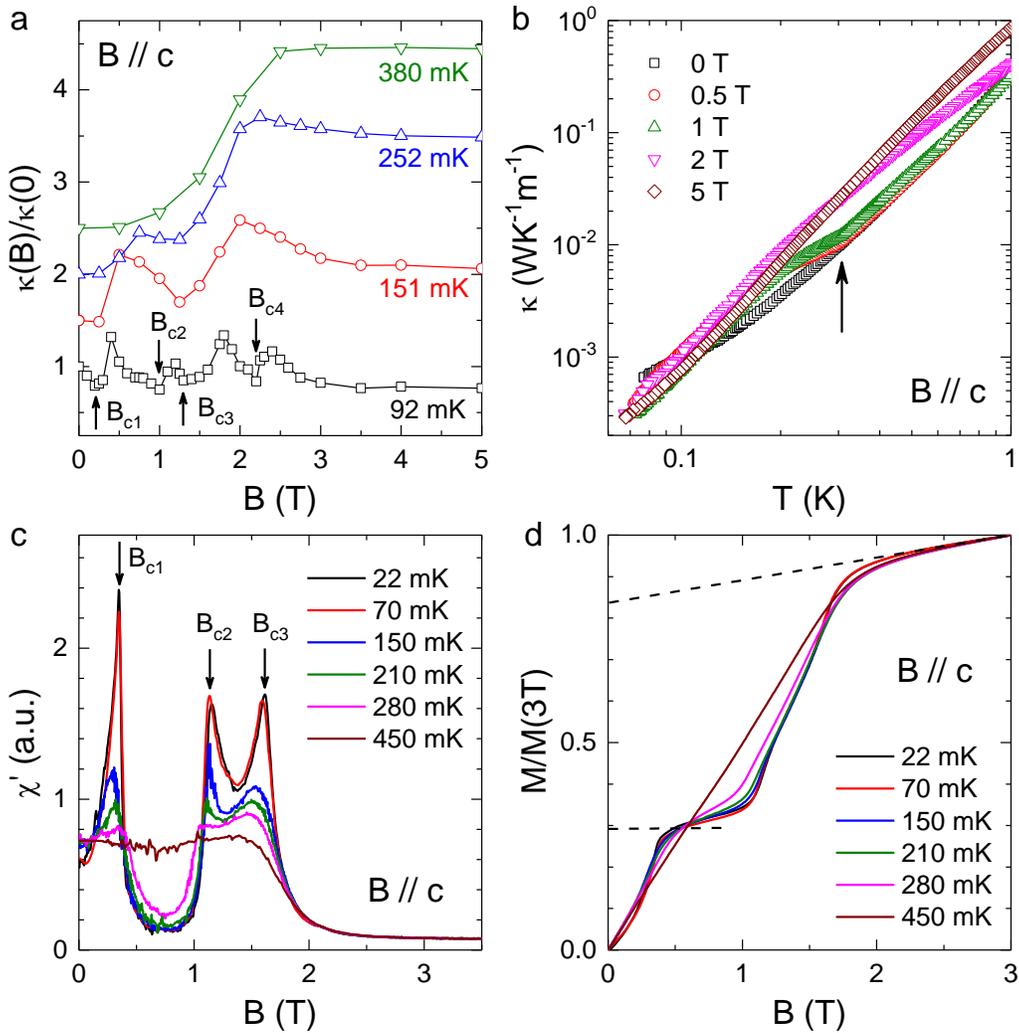

**Figure 4  Thermal conductivity and AC susceptibility of $Na_2BaCo(PO_4)_2$ for *B // c*. a**, Magnetic-field dependence of thermal conductivity at different temperatures. For clarifying, the 151 mK, 252 mK and 380 mK curves are shifted upward by 0.5, 1.0 and 1.5, respectively. At 92 mK, there are four minima indicated by arrows. With increasing temperature, the minima become weaker and disappear above 380 mK. **b,** Temperature dependence of $\kappa$ in different fields. At low fields of 0.5 and 1 T, there is a clear slope change of $\kappa(T)$ curves around 310 mK, which has good correspondence to the specific-heat anomaly. **c,** AC magnetic susceptibility at different temperatures. There are three peaks in the low temperature curves. **d,** Magnetization curves obtained by integrating the AC susceptibility data and renormalized with the 3 T value. The lower dashed line indicates a ~1/3 plateau and the upper dashed line indicates the Van Vleck paramagnetic background.

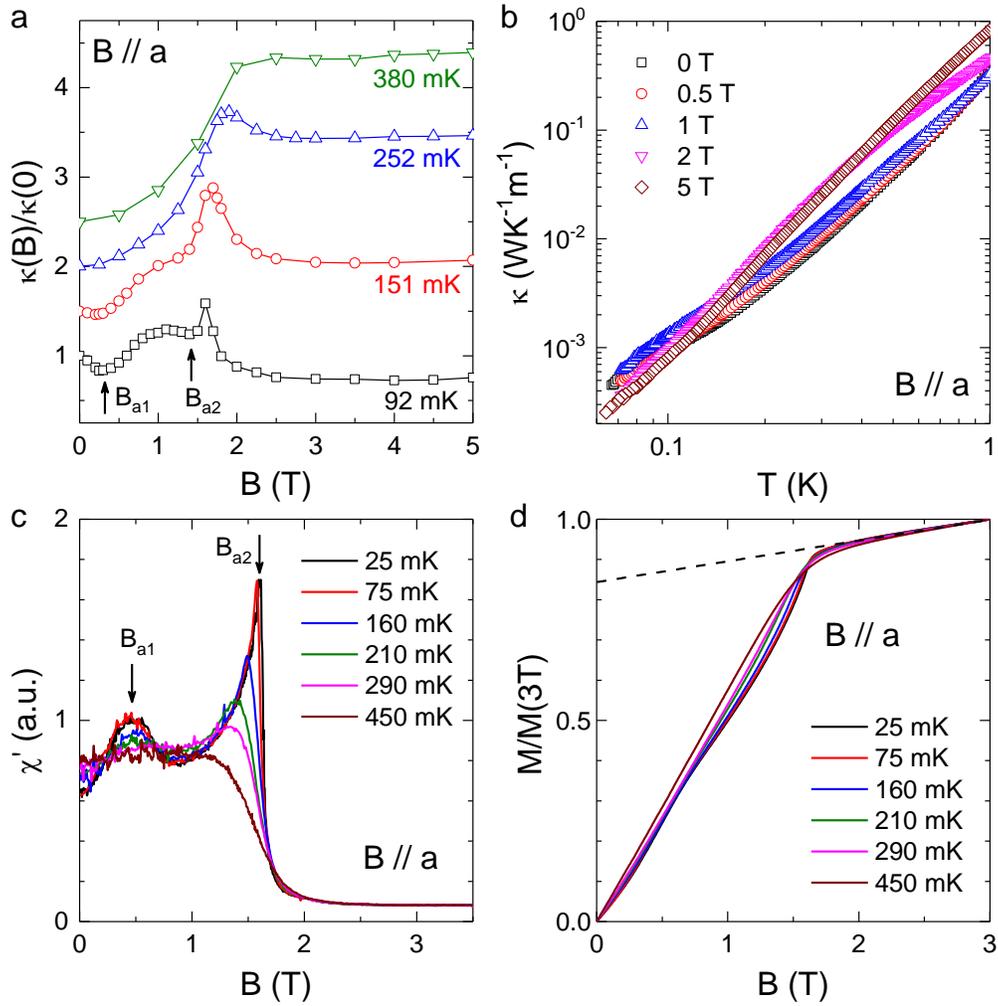

**Figure 5  Thermal conductivity and AC susceptibility of $Na_2BaCo(PO_4)_2$ for $B$ // $a$. a**, Magnetic-field dependence of thermal conductivity at different temperatures. For clarifying, the 151 mK, 252 mK and 380 mK curves are shifted upward by 0.5, 1.0 and 1.5, respectively. At very low temperatures, there are two minima on the $\kappa(B)$ curves which disappear above 252 mK. **b,** Temperature dependence of $\kappa$ in different fields. No clear anomaly is observed. **c,** AC magnetic susceptibility at different temperatures. There are two peaks in the low temperature curves. **d,** Magnetization curves obtained by integrating the AC susceptibility data and renormalized with the 3T value. The dashed line indicates the Van Vleck paramagnetic background.

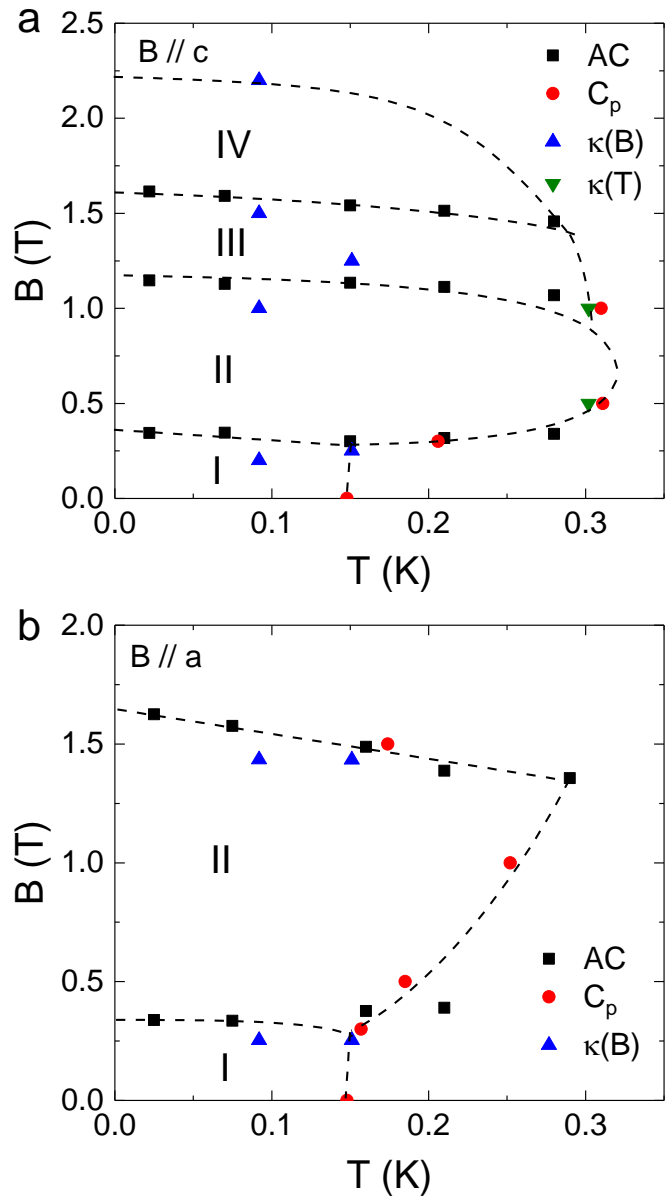

**Figure 6 The magnetic phase diagrams of Na$_2$BaCo(PO$_4$)$_2$ for *B // c* (a) and *B // a* (b).** The data points are obtained from the AC susceptibility (AC), specific heat ($C_p$) and temperature or field dependence of thermal conductivity ($\kappa(T)$ and $\kappa(B)$) measurements. The dashed lines are phase boundaries. For *B // c*, there are four phases (I, II, III, and IV) in the low-temperature and low-field region. Whereas, there are two phase (I and II) at low-temperature and low-field for *B // a*. The dashed lines are a guide to the eye.

**Methods**

**Sample preparation and characterization.** The single crystals were grown by flux method as reported in ref. 46. One adjustment made is that Platinum crucibles instead of alumina crucibles were used in our growth. The powder X-ray diffraction measurement on the ground single crystals confirmed its lattice structure is the same as reported in ref. 46. Laue back diffraction confirmed the flat surface of the as grown crystals is the *ab* plane. DC magnetic susceptibility was measured with a Quantum Design superconducting quantum interference device (SQUID) magnetometer. The applied field $B = 0.1$ T is parallel to the *ab* plane. Specific heat was measured with a Quantum Design Physical Property Measurements System (PPMS), equipped with a dilution refrigerator insert.

**AC susceptibility measurements.** The AC susceptibility measurements were conducted with a voltage controlled current source (Stanford Research, CS580) and lock-in amplifier (Stanford Research, SR830). The phase of the lock-in amplifier is set to measure the first harmonic signal. Single crystal samples of $Na_2BaCo(PO_4)_2$ were prepared to allow the AC and DC magnetic fields to be perpendicular and parallel to the *c* axis separately in the measurements. The rms amplitude of the ac excitation field is set to be 0.6 Oe with the frequency fixed to be 220 Hz. The measurements were performed at SCM1 of the National High Magnetic Field Laboratory, Tallahassee, by using a dilution refrigerator. The data was obtained by the zero field cooling process and we increased the magnetic field during the ramping process.

**Thermal conductivity measurements.** Thermal conductivity was measured by using a "one heater, two thermometers" technique in a $^3$He/$^4$He dilution refrigerator and in a $^3$He refrigerator, equipped with a 14 T superconducting magnet[63-73]. The sample was cut precisely along the crystallographic axes with dimensions of $3.0 \times 0.63 \times 0.14$ mm$^3$, where the longest and the shortest dimensions are along the *a* and the *c* axes, respectively. The heat currents were applied along the *a* axis while the magnetic fields were applied along either the *a* or *c* axis. Since the AC susceptibility clearly showed no hysteresis with sweeping field, we did not perform all the specific heat and thermal conductivity measurements with the zero-field cooling process. However, we carefully checked the first $\kappa(B)$ measurement at 92 mK for both *B* // *c* and *B* // *a*. The sample was zero-field cooled to 92 mK and the $\kappa$ was measured with increasing field to 14 T and then decreasing field to 0 T. No hysteresis was observed in $\kappa(B)$.

For low-temperature thermal conductivity measurements, calibrating the magnetoresistance effect of resistor thermometers is a basic requirement. The thermometers (RuO$_2$) used at 300m K – 30 K in the $^3$He refrigerator are pre-calibrated by using a capacitance sensor (Lakeshore Cryotronics) as the reference[65,67-69]; the thermometers (RuO$_2$) used at 50 mK – 1 K in the dilution refrigerator are pre-calibrated by using a RuO$_2$ reference sensor (Lakeshore Cryotronics) mounted at the mixture chamber (the superconducting magnet was equipped with a cancellation coil at the height of mixture chamber)[70-72]. The resolution of the $\kappa$ measurements is typically better than 3% (better at higher temperature). The sample size was determined by using microscopy and has uncertainty of < 5%. Therefore, the total error bar of is $\kappa$ always < 8%. The uncertainty of $\kappa_0/T$ caused by the fitting is about 2%. The $\kappa_0/T$ value of Na$_2$BaCo(PO$_4$)$_2$ is ~ 30 times smaller than that of EtMe$_3$Sb[Pd(dmit)$_2$]$_2$ and ~10 times smaller than that of 1T-TaS$_2$[18-20]. One may ask whether this value is too small to be resolved by the $\kappa$ measurement at ultralow temperatures. We would like to mention that this residual thermal conductivity is actually comparable to those in high-$T_c$ cuprate superconductors (HTSC). For HTSC, the $\kappa_0/T$ is contributed by the nodal quasiparticles from the $d$-wave superconducting state and has been experimentally observed by us in many materials, including La$_{2-x}$Sr$_x$CuO$_4$, YBaCu$_3$O$_y$, Bi$_2$Sr$_{2-x}$La$_x$CuO$_{6+\delta}$, and Bi$_2$Sr$_2$CaCu$_2$O$_{8+\delta}$[64-67]. In these materials, it is well resolved that the $\kappa_0/T$ varies from 0.0015 to 0.06 WK$^{-2}$m$^{-1}$ and shows systematic changes with the carrier concentration. Similar experimental results have also been reported by other groups for both the cuprate and the iron-based superconductors[74-77]. Therefore, a $\kappa_0/T$ value of 0.0062 WK$^{-2}$m$^{-1}$ is big enough to be correctly detected by a high-level measurement.

**Demagnetization effect.** Here we list the used samples' dimensions and weights for various measurements. AC susceptibility: for $B // a$, 1.44 × 1.15 × 4.80 mm$^3$, 33.2 mg; for $B // c$, 1.30 × 1.30 × 5.20 mm$^3$, 36.7 mg. For both cases, the field is along the longest dimension. Specific heat: for $B // a$, 1.96 × 0.45 × 0.38 mm$^3$, 1.40 mg; for $B // c$, 1.87 × 1.16 × 0.18 mm$^3$, 1.63 mg. For both cases, the field is along the shortest dimension. Thermal conductivity: 3.0 × 0.63 × 0.14 mm$^3$, 1.11 mg. For $B // a$, the field is along the longest dimension, for $B // c$, the field is along the shortest dimension. The estimated upper limit of the modification of $B$ by the demagnetization effects for AC susceptibility is less than 1 % for both directions, for specific heat is less than 4 % for *B // a* and less than 8 % for *B // c*, and for thermal conductivity is less than 1 % for *B // a* and less than 8 % for *B // c*. Such kind of small modification was neglected.

**Data availability**

The data that support the findings of this study are available from the corresponding author upon reasonable request.

**Acknowledgements**

The work at the University of Science and Technology of China (N.L., W.J.C., X.Z. and X.F.S.) was supported by the National Natural Science Foundation of China (Grants No. U1832209 and No. 11874336), the National Basic Research Program of China (Grant No. 2016YFA0300103), the Innovative Program of Hefei Science Center CAS (Grant No. 2019HSC-CIP001), and the Users with Excellence Project of Hefei Science Center CAS (Grant No. 2018HSC-UE012). The work at the University of Tennessee (Q.H., Q.C. and H.D.Z.) was supported by the National Science Foundation through award DMR-2003117. A portion of this work was performed at the National High Magnetic Field Laboratory, which is supported by the National Science Foundation Cooperative Agreement No. DMR-1644779 and the State of Florida. Q.H. and H.D.Z. are very grateful to Ruidan Zhong for her helpful discussion on sample synthesis.

**Author contributions**

N.L. and W.J.C. and X.F.S. performed thermal conductivity measurements and analyzed the data with help from X.Z. and H.D.Z. X.Y.Y. performed the specific heat measurements. Q.H., Q.C., E.S.C., and H.D.Z. made the samples and performed the low-temperature AC susceptibility measurements. X.F.S. and H.D.Z. wrote the paper with input from all other co-authors.

**Competing financial interests**

The authors declare no competing interests.

**Additional Information**

**Correspondence and requests for materials** should be addressed to X.F.S. or H.D.Z.

# Supplementary Information for
## *Possible itinerant gapless excitations and quantum spin state transitions in the effective spin-1/2 triangular-lattice antiferromagnet $Na_2BaCo(PO_4)_2$*

N. Li[1,6], Q. Huang[2,6], X. Y. Yue[3,6], W. J. Chu[1], Q. Chen[2], E. S. Choi[4], X. Zhao[5], H. D. Zhou[2]★, and X. F. Sun[1,3]★

[1]Department of Physics, Hefei National Laboratory for Physical Sciences at Microscale, and Key Laboratory of Strongly-Coupled Quantum Matter Physics (CAS), University of Science and Technology of China, Hefei, Anhui 230026, People's Republic of China

[2]Department of Physics and Astronomy, University of Tennessee, Knoxville, Tennessee 37996-1200, USA

[3]Institute of Physical Science and Information Technology, Anhui University, Hefei, Anhui 230601, People's Republic of China

[4]National High Magnetic Field Laboratory, Florida State University, Tallahassee, FL 32310-3706, USA

[5]School of Physical Sciences, University of Science and Technology of China, Hefei, Anhui 230026, People's Republic of China

[6]These authors contributed equally: N. Li, Q. Huang, X. Y. Yue.

★email: hzhou10@utk.edu; xfsun@ustc.edu.cn

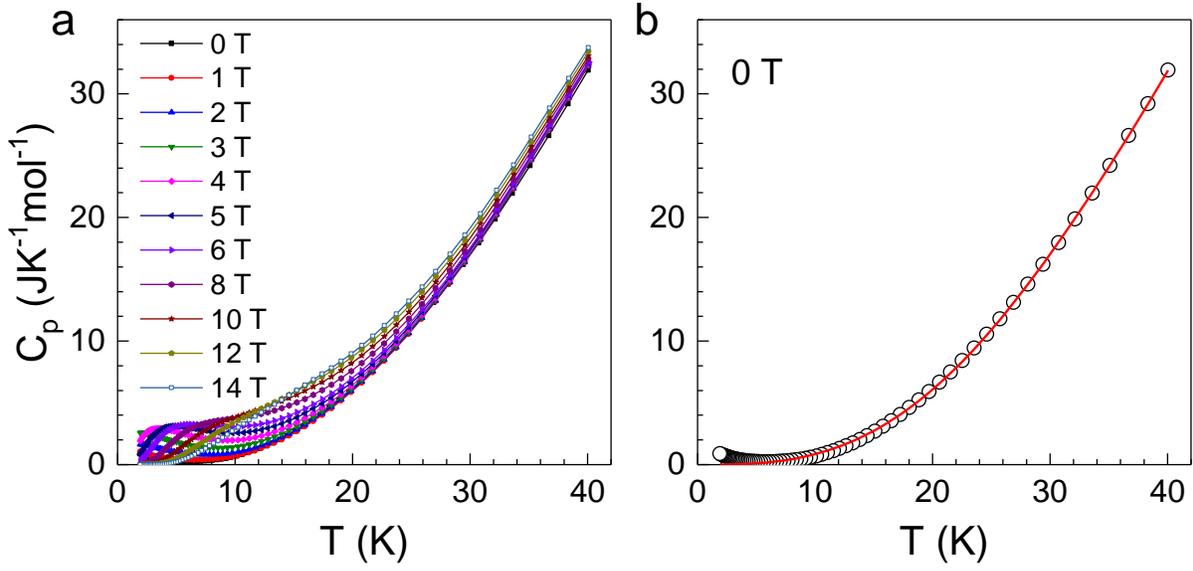

**Supplementary Figure 1** a, Specific heat of $Na_2BaCo(PO_4)_2$ single crystal at 1.9 – 40 K and in various magnetic fields up to 14 T. b, Zero-field data. The solid line shows the fitting to data at $T >$ 10 K by using the formula of phonon specific heat, $C_{ph} = \beta T^3 + \beta_5 T^5 + \beta_7 T^7$.

Figure 1a shows the specific heat data of $Na_2BaCo(PO_4)_2$ single crystal at 1.9 – 40 K and in various magnetic fields of 0 – 14 T. These data are consistent with those reported in by Zhong *et al.*[1] There is no sign of phase transition in this temperature range. With applying magnetic field, a broad peak appears at low temperature. To estimate the phonon specific heat, we fit the zero-field data at 10 – 40 K by using the low-frequency expansion of the Debye function, $C_{ph} = \beta T^3 + \beta_5 T^5 + \beta_7 T^7$, where $\beta$, $\beta_5$, and $\beta_7$ are temperature-independent coefficients[2], as shown by the solid line in Fig. 1b. The fitting parameters are $\beta = 8.83 \times 10^{-4}$ $JK^{-4}mol^{-1}$, $\beta_5 = -3.32 \times 10^{-7}$ $JK^{-6}mol^{-1}$, and $\beta_7 = 6.67 \times 10^{-11}$ $JK^{-8}mol^{-1}$. Note that at very low temperatures, the $T^5$ and $T^7$ terms are negligible and the phonon specific heat shows a well-known $T^3$ dependence with the coefficient of $\beta$. As also mentioned in the main text, the phonon velocity can be calculated from the $\beta$ value as $v_{ph} = 2430$ $ms^{-1}$, which is of a reasonable value for oxide materials. For a comparison, the phonon velocity of another QSL candidate, $Tb_2Ti_2O_7$, is 3440 $ms^{-1}$ [3], determined in the same method.

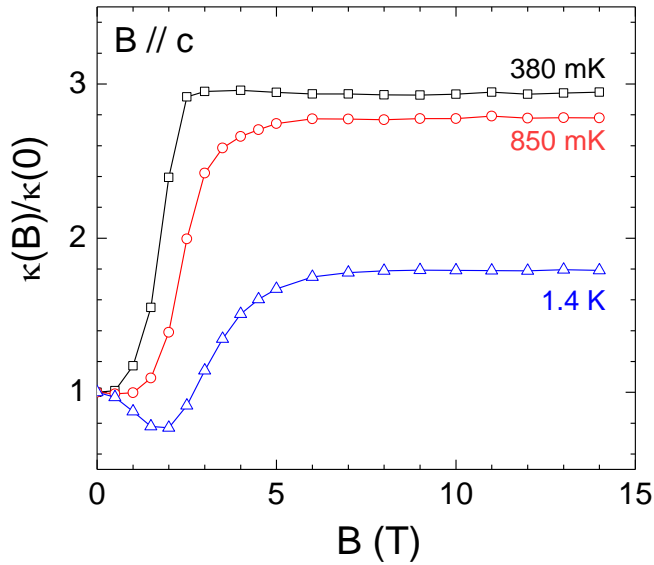

**Supplementary Figure 2**   Magnetic field dependence of thermal conductivity for $Na_2BaCo(PO_4)_2$ single crystal at different temperatures with $B // c$.

Figure 2 shows magnetic field dependence of thermal conductivity for $Na_2BaCo(PO_4)_2$ single crystal at 380 mK, 850mK and 1.4 K. At higher temperature, the $κ(B)$ shows a bit more different behavior, that is, with increasing field the $κ$ first decreases at low field and then increases. This phenomenon is similar to that observed in some other magnetic systems, like $Yb_2Ti_2O_7$[4], and can be understood as the opposite roles of spin excitations as both carrying heat and scattering phonons.